\documentclass[aps,pra,twocolumn,showpacs]{revtex4}
\usepackage{amsbsy,latexsym}
\usepackage{amsfonts}
\usepackage{amssymb}
\usepackage[mathscr]{eucal}

\newcommand{\ket}[1]{\vert #1 \rangle}
\newcommand{\bra}[1]{\langle #1 \vert}

\newcommand{\D}[3]{\mathfrak D^{(#1)}_{#2 #3}}
\newcommand{\Dc}[3]{\mathfrak D^{(#1) *}_{#2 #3}}

\newcommand{\threeJ}[6]{\mbox{\footnotesize$
 \displaystyle\pmatrix{#1&#2&#3\cr #4&#5&#6}$}}
 \newcommand{\threeJt}[6]{\mbox{\tiny$
 \displaystyle\pmatrix{#1&#2&#3\cr #4&#5&#6}$}}
\begin{document}
\title{Communication of Spin Directions with
Product States and Finite Measurements}
\author{E.~Bagan, M.~Baig, and R.~Mu{\~n}oz-Tapia}
\affiliation{Grup de F{\'\i}sica Te{\`o}rica \& IFAE, Facultat de
Ci{\`e}ncies, Edifici Cn, Universitat Aut{\`o}noma de Barcelona, 08193
Bellaterra (Barcelona) Spain}
%\date{\today}

\begin{abstract}
Total spin eigenstates  can be used to intrinsically encode a
direction, which can later be decoded by means of a quantum
measurement. We study the optimal strategy that can be adopted if,
as is likely in practical applications, only product states of
$N$-spins are available. We obtain the asymptotic behaviour of the
average fidelity which provides a proof that the optimal states
must be entangled. We also give a prescription for constructing
finite measurements for general encoding eigenstates.

\end{abstract}

\maketitle

%\section{Introduction}\label{sect-introduction}
Quantum Mechanics is rapidly broadening our knowledge of the ways
information can be stored, transmitted and retrieved.  Here we
address the concrete issue of communicating information of a
direction using quantum states, which has attracted  much
attention \cite{mp,derka,lpt,gp,massar,us,ps,us2}. Consider two
parties, Alice and Bob, and imagine that Bob is lost in space and
Alice wants to tell him the direction home. If communication by
standard means is difficult, she can encode the direction in a
quantum system and physically send it to Bob. Alice's quantum
state must {\em intrinsically} point along the direction, given
by the unit vector $\vec{n}$. If we assume that her system is
made out of $N$ spins, then it must be an eigenstate of
$\vec{n}\cdot\vec{S}$, where $\vec S$ is the total spin
\cite{referee}. After he receives the state, Bob can  perform a
quantum measurement and retrieve Alice's direction with some
accuracy. {}From each outcome (labeled with an index~$r$) of the
measurement, Bob will guess a direction, given by a unit
vector~$\vec n_{r}$. We use the fidelity, $(1+\vec n\cdot\vec
n_{r})/ 2$, as a figure of merit (we have also computed the
information gain for a check of our conclusions). An average
fidelity $F=1$ means a perfect determination of the direction. We
can view $N$ as the size of the resources available to Alice.
Obviously, the average fidelity should increase as the resources
increase. However, for a given number of resources, the actual
value of the average fidelity and the rate it approaches to one
depend on the type of states being used. For instance, the
maximal average fidelity (MAF) for states of~$N$ parallel spins,
$\ket{\uparrow\uparrow\cdots \uparrow}$, is
$F=(N+1)/(N+2)$~\cite{mp}, which is readily seen to approach
unity linearly: $F\sim 1-1/N$. If the resources consist of only
two spins, choosing them to be antiparallel,
$\ket{\uparrow\downarrow},\ket{\uparrow\downarrow}$,  leads to a
value of~\cite{gp} $F=(3+\sqrt{3})/6$, which is larger than
$F=3/4$ for two parallel spins. Thus, one can improve on the
communication of a direction without increasing the resources.
In~\cite{us,us2} we generalized these results to arbitrary~$N$
and computed the MAF optimizing both Alice's states and Bob's
measurements. We proved that the MAF approaches unity as $F\sim
1- 5.8/N^2$, i.e, there is a quadratic improvement on the quality
of the communication process over the parallel case. The optimal
states that lead to such MAF are, in fact, a whole family of
states which for $N>2$ does not seem to include any  state of the
form $\ket{\uparrow\downarrow\downarrow\cdots\uparrow}$ (we will
loosely refer to these states  as product states). {}From the
practical point of view, however, product states are of crucial
importance, since they are likely to be the only ones that can be
used in real devices (although they are expected not to be
optimal). There are then two obvious questions one would like to
answer. Firstly, among these states, what are the best for
encoding a direction? And, secondly, is there a quadratic
improvement in the rate the MAF approaches to one?  We will
answer these questions in this paper. We show that the optimal
product states are those with the smallest $|m|$, where $m$ is the
eigenvalue of~$\vec n\cdot\vec S$, and that the corresponding MAF
for large $N$ is $F\sim 1-1/(2N)$. This result proves our implicit
assumption that the truly optimal states are entangled for $N>2$.
Although product states do not exhibit the quadratic behaviour in
$1/N$ of the truly optimal ones, we see that they are still much
better than the $N$ parallel spin states for communicating a spin
direction.

To compute the MAF of an optimal measurement, it is useful to
consider a positive operator valued measurement (POVM) with
infinitely many outcomes or continuous POVM~\cite{holevo}.
We show, however, that
one can always construct optimal POVMs with a finite number of
outcomes. This is an important point since these are the only
measurements that can be physically implemented. For parallel
encodings, there are explicit realizations of optimal finite
POVMs for arbitrary~$N$~\cite{derka} and minimal versions of
these for~$N\leq 7$ can be found in~\cite{lpt}. The outcomes of
these POVMs are associated with unit vectors $\vec n_{r}$
(directions) that we can picture as the vertices of certain
polyhedra inscribed in the unit sphere. In this paper we prove
that the very same polyhedra define optimal measurements for very
general encoding states  and that the minimal polyhedra
of~\cite{lpt} remain minimal for these general states.
% We will
%closely follow the notation and conventions of~\cite{us2}.

 Alice's
states can be obtained by rotating a fixed eigenstate
of~$S_z=\vec{z}\cdot\vec{S}$ that we denote by~$\ket{A}$. In
terms of the individual spins it is just of the form
$\ket{\uparrow\downarrow\downarrow\uparrow\cdots}$. It is
convenient to write all quantum states in terms of the
irreducible representations of $\rm SU(2)$, thus
\begin{equation}\label{A-general}
  \ket{A}=\sum_{j=m}^{N/2}\left(\sum_\alpha A^{\alpha}_j
\ket{j,m;\alpha}\right),
\end{equation}
where  the first two labels are the usual quantum numbers of the
total spin $\vec{S}^2$ and its third component $S_z$, i.e,
$\vec{S}{}^2\ket{j,m;\alpha}=j(j+1)\ket{j,m;\alpha}$ and $ S_z
\ket{j,m;\alpha} =m\ket{j,m;\alpha}$. The third index, $\alpha$,
labels different occurrences of the same
representation~$\boldsymbol{j}$ in the Clebsch-Gordan
decomposition of $(\mbox{{\boldmath
$1/2$}})^{\mbox{\scriptsize{\boldmath $\otimes$}} N}$. Also
from~\cite{us2}, one can show that there exists an optimal
continuous POVM, defined by a complete set of positive projectors
of the form $O(\vec n)=U(\vec n)\left[\ket{B}\bra{B}+
\ket{B'}\bra{B'}+\cdots\right]U^\dagger(\vec n)$, where $U(\vec
n)$ is the element of $\rm SU(2)$ associated with the rotation $R:\vec
z \mapsto \vec n$, and $\ket{B}$, $\ket{B'}$,~\dots, are fixed
states given by linear combinations entirely analogous
to~(\ref{A-general}). The average fidelity is
\begin{equation}
F=\int dn{1+\vec{z}\cdot\vec{n}\over 2} \bra{A}O(\vec{n})\ket{A}.
\label{fidelity}
\end{equation}

To compute  (\ref{fidelity}) one can use just the effective state
$\ket{\tilde B}=\sum_{j=m}^{N/2}  \sqrt{2j+1}\ket{j,m}$, instead
of employing all of $\ket{B}$, $\ket{B'}$,\dots, i.e., $O(\vec
n)\to U(\vec n) \ket{\tilde B}\bra{\tilde B}U^\dagger(\vec n)$.
Similarly, for given quantum numbers $j$, $m$, we  define the
effective components of~$\ket{A}$  as
%\begin{equation}\label{Aj}
$  \tilde{A}_{j}\equiv
    \sqrt{\sum_{\alpha}(A_{j}^{\alpha})^2}$,
%\end{equation}
which contains the information required to compute the MAF. For
any $\ket{A}$ of the form $\ket{\uparrow\downarrow\uparrow\cdots}$
with $n_\uparrow$ spins up and $n_{\downarrow}$ spins down, the
MAF in (\ref{fidelity}) can be computed using the effective state
$\ket{\tilde A}=\sum_{j=m}^{N/2}\tilde{A}_j \ket{j,m}$, where
$m=(n_{\uparrow}-n_{\downarrow})/2$ and the coefficients
$\tilde{A}_j$ are explicitly given by
\begin{equation}\label{Aj-2}
 \tilde{A}_j=\sqrt{\frac{1+2j}{J+1+j}}\sqrt{\frac{(J-m)!(J+m)!}{(J-j)!(J+j)!}};
          \quad J\equiv {N\over2}.
\end{equation}
We obtain the following MAF:
\begin{equation}\label{fidelity-2}
F=\frac{1}{2}+\frac{1}{2}\sum_{j=m}^{J}  \mu_j \tilde{A}_j^2 +
\sum_{j=m+1}^{J}  \tilde{A}_{j-1} \tilde{A}_{j} \nu_j,
\end{equation}
where \cite{us2}
%
%\begin{equation}
$\mu_j={m^2}/\;{j(j+1)}$,
$\nu_j={j(j^2-m^2)}/{\sqrt{4 j^2-1}} \label{muj-nuj}$.
%\end{equation}
%

We have written equal quantum numbers $m$ for $\ket{A}$ and
$\ket{B}$. Note that if  $m_B> m_A$,  $O(\vec n)$ would not be a
complete set of projectors on the whole Hilbert space spanned by
$U(\vec n)\ket{A}$; conversely, if  $m_B< m_A$,  Alice's states do
not use the full capabilities of Bob's measuring device and the
strategy cannot be optimal.

The maximal fidelity in (\ref{fidelity-2}) is attained for the
minimal value of $|m|$ (this is $m=0$ for $N$ even and $m= 1/2$
for $N$ odd), i.e., for maximal antiparallel spins. In Table~I we
collect the values of the MAF for up to~$N=7$ and we compare them
with the MAFs of parallel ($F_{P}$)~\cite{mp} and optimal
($F_{O}$)~\cite{us2} encodings. Note that antiparallel product
states lead to MAFs ($F_{A}$) remarkably close to the optimal
ones. Moreover, one can easily prove that antiparallel spins are
better than parallel ones for encoding a direction. We now show
this for an even number of spins, $N=2n$, and $m=0$, in which
case the MAF (\ref{fidelity-2}) takes the simple form
\begin{equation}\label{fidelity-3}
F_{A}=\frac{1}{2}+\sum_{j=1}^{n} \frac{n!^2}{(n-j)! (n+j)!}
\frac{j}{\sqrt{(n+1)^2-j^2}}.
\end{equation}
Setting $j=0$ inside the square root, we obtain
\begin{equation}\label{max}
  F_{A}>\frac{1}{2}+\sum_{j=1}^{n} {n!^2\,j/(n+1)\over(n-j)! (n+j)!}
=\frac{N+1}{N+2}=F_{P}.
\end{equation}

\begin{table}\label{table-I}
\begin{center}
\begin{tabular}{c|cccccc}
  \toprule
  $N$  & 2 & 3 & 4 & 5 & 6 & 7 \\
  \colrule
$F_P$ & 0.75   & 0.8    & 0.8333 & 0.8571 & 0.875  & 0.8889
\\
$F_A$ & 0.7887 & 0.8444 & 0.8848 & 0.9069 & 0.9235 & 0.9342
\\
$F_O$ & 0.7887 & 0.8449 & 0.8873 & 0.9114 & 0.9306 & 0.9429
\\
   \colrule
$I_P$ & 0.6232 & 0.9180 & 1.1678 & 1.3827 & 1.5708 & 1.7376
\\
$I_A$ & 0.8664 & 1.2816 & 1.7077 & 2.0079 & 2.2873 & 2.4897
\\
$I_O$ & 0.8664 & 1.2925 & 1.7589 & 2.1086 & 2.4685 & 2.7548
\\
  \botrule
\end{tabular}
\end{center}
\caption{Maximal average fidelities ($F$) and information gains
($I$) for parallel ($P$), antiparallel ($A$) and optimal ($O$)
encodings}
\end{table}
%%%%%%%%%%%%%%%%%%%%%%%%%%%%%%%%%%%%%%%%%%%%%%%%%%%%%%%%%%%%%%%%%%%%%%%%%%%%%%%%

We would next like to study the large $N$ asymptotic behaviour of
$F_{A}$ to see whether it exhibits the quadratic behaviour of the
optimal states $1-F_O\sim 1/N^2$. We just have to
compute~(\ref{fidelity-3}) for large $n$. Notice first that, using
the Stirling approximation, we have the following limit
\begin{equation}\label{limit}
  \frac{n!^2}{(n-j)! (n+j)!}\to e^{-j^2/n
  }\left(1+\frac{j^2}{2 n^2}-\frac{j^4}{6 n^3}
  +\cdots\right).
\end{equation}
Therefore, only terms with $j\sim \sqrt{n}$ give a significant
contribution to the sum in~(\ref{fidelity-3}). Hence, it is
legitimate to expand the square root in~(\ref{fidelity-3}) in
powers of $j$. The resulting expression can be evaluated by means
of the Euler-Maclaurin formula \cite{ayant}
\begin{eqnarray}
 \sum_{j=1}^{n} {1\over n}f(j/n) &=&\int_0^1 dx f(x)+
{f(1)-f(0)\over 2n}\nonumber \\
&+&{f'(1)-f'(0)\over12n^2}- \cdots, \label{euler-mac}
\end{eqnarray}
where in our case $f(j/n)$ is the product of the right hand side
of~(\ref{limit}) times the expansion of $j/\sqrt{(n+1)^2-j^2}$.
Taking into account all the relevant terms, one obtains that up
to order $1/n$,
\begin{equation}\label{fidelity-large-n}
  F_{A}=1-\frac{1}{4n}+\cdots=1-\frac{1}{2N}+ \cdots .
\end{equation}
Therefore, antiparallel spin states lead to a MAF that approaches
unity in $1/N$, faster than it does for parallel spins, but only
because of the smaller negative coefficient of the $1/N$ term
($1/2$ compared to $1$). In this sense, both types of encodings
are qualitatively similar. The quadratic behaviour of truly
optimal states (which are entangled) cannot be attained by any
product state. It is lengthier, but straightforward, to compute
the subleading term in~(\ref{fidelity-large-n}). We obtain the
following compact expression for the MAF:
\begin{equation}\label{fidelity-N}
 F_{A}=\frac{2N+1}{2N+2}+O(1/N^3).
\end{equation}

To check that our results are not an artifact~of~our
par\-ti\-cular fi\-gu\-re of merit, we have also computed the
average information gain~\cite{peres}, $I=\int d n \;\bra{A}
O(\vec{n}) \ket{A}\log _{2}\left(\bra{A} O(\vec{n}) \ket{A}
\right)$, for parallel, antiparallel and optimal states. Our
results are also collected in Table~I. We see that both,
information gain and fidelity, exhibit the same pattern. Namely,
optimal (entangled) states lead to the largest $I$ and $F$, but
antiparallel spins  have values very close to the optimal ones and
much larger than those of parallel spins.

Up to now we have dealt with continuous POVMs. They are useful
mathematical tools that simplify the calculation of the MAF for
any optimal measurement on an isotropic distribution of
directions. The projectors $O(\vec n)$ satisfy the closure
relation $\int dn\,O(\vec n)=\mathbb I$ because the orthogonality
of the  non-equivalent irreducible  $\rm SU(2)$ representations, $\D j
m {m'}$,  under the isotropic integration over the unit sphere.
However, only  POVMs with a finite number of outcomes can be
realized in nature. Unfortunately, finite POVMs are rather
elusive because there is no clear and unique definition of
isotropy for a finite set of directions (unit vectors) $\vec
n_{r}$. We provide here a functional definition, which will
enable us to give a general algorithm for constructing optimal
and finite POVMs. Moreover, it will become obvious that the
problem of discretizing a POVM is of geometrical nature.

In the context of this paper, we say that a finite set  of unit
vectors ${\vec n_{r}}$ is isotropically distributed up to spin $J$
if there exist positive weights $\{c_{r}\}$ such that the
following orthogonality relation holds for any $j,j'\leq J$:
\begin{equation}
\sum_{r=1}^{N(J)}  c_{r}\,\D j m k (\vec n_{r}) \Dc {j'} {m'} k
(\vec n_{r})
 =\frac{C_J}{2j+1}\delta_m^{m'}\delta_{j}^{j'},
 \label{eb1f}
\end{equation}
where $C_J=\sum_{r=1}^{N(J)} c_r$ is the equivalent of the solid
angle $4\pi$ in the continuous orthogonality relation $
 \int d\Omega\; \D j m k (\vec n) \Dc {j'} {m'} k  (\vec n)
 ={4\pi} \delta_m^{m'}\delta_{j}^{j'}/ (2j+1),
$ and $N(J)$ is the number of elements of $\{\vec n_{r}\}$. Here
we use the shorthand notation $\D j m k(\vec n)=\D j m
k(\phi,\theta,0)$, where $\vec
n=(\sin\theta\cos\phi,\sin\theta\sin\phi,\cos\theta)$, and
$\alpha$, $\beta$, $\gamma$  in $ \D j m k(\alpha,\beta,\gamma)$
are the standard Euler angles. The main difference between the
continuous orthogonality relation and~(\ref{eb1f}) is that the
latter can only hold for $j$, $j'$ up to a maximal value $J$. The
larger $J$ is, the larger $N(J)$ must be chosen.

We will now show that~(\ref{eb1f}) is equivalent to %the conditions
\begin{equation}
 \sum_{r=1}^{N(J)}  c_{r}\,Y^{M}_{L}(\theta_{r},\phi_{r})=0;\qquad
 \begin{array}{rcl}L&=&1,2,\dots,2J;\\ M&=&0,1,\dots, L;
 \end{array}
 \label{eb2}
\end{equation}
where $Y^m_{l}(\theta,\phi)$ are the standard spherical harmonics.
Eq.~\ref{eb2} is very appealing since one can establish a physical
analogy. If we view $c_{r}$ as a (positive) charge at the
position~$\vec n_{r}$, Eq.~\ref{eb2} tell us that~(\ref{eb1f}) is
equivalent to the requirement that electrostatic multipoles of
order less or equal to $2J$ vanish. The conditions~(\ref{eb2}) are
exactly those given in~\cite{lpt} for minimal and optimal POVMs in
the case of a signal state consisting of~$N$ parallel spins. We
see here that (\ref{eb2}) are actually of much greater generality.
To simplify the notation it is convenient to define the
quantities:
\[
  z_L^M=\sum_r^{N(J)}c_r \Dc L M 0 (\vec{n}_r)
  =\frac{\sqrt{4\pi}}{\sqrt{2L+1}}
 \sum_{r}^{N(J)}
   c_{r}\,Y^{-M}_{L}(\theta_{r},\phi_{r}).
\]
Now (\ref{eb2}) simply reads  $z_L^M=0$ for all $L$ and $M$ listed
there. In the following, $j$ and $j'$ are required to satisfy
$j,j'\leq J$. The group theoretical results that will be used
below are mainly borrowed from \cite{edmonds}. Note first that
the product $\D {j}{m}{k} \Dc {j'}{m'}{k}$ in (\ref{eb1f}) can be
written as a sum of $\Dc{l} {m-m'}{0} \propto z_l^{m-m'}$.
Explicitly, (\ref{eb1f}) is equivalent to the set of linear
equations
\begin{eqnarray}
 &&\sum_{l}(2l+1)\threeJ{j}{j'}{l}{m}{-m'}{m-m'}
 \threeJ{j}{j'}{l}{k}{-k}{0} z_l^{m-m'}\nonumber \\
 &&=(-1)^{m'-k}{C_J
\over 2j+1} \delta_m^{m'}\delta_{j}^{j'},
 \label{eb3}
\end{eqnarray}
where $\threeJt{j}{j'}{j''}{m}{m'}{m''}$ are the 3-j symbols and
the sum runs over all $l$ satisfying the triangular condition (in
particular $l\le 2J$). By direct substitution it is trivial to
check that~(\ref{eb2}) is a solution of~(\ref{eb3}) for all
relevant $j,j'$ and  $m,m'$. Therefore~(\ref{eb2}) are sufficient
conditions. To prove that~(\ref{eb2}) are also necessary, we
multiply~(\ref{eb3}) by $\threeJt{j}{j'}{L}{m}{-m'}{M}$ and sum
over $m$ and $m'$. Next, we use the orthogonality
condition~\cite{edmonds}
\begin{equation}
 \sum_{mm'}\threeJ{j}{j'}{l}{m}{m'}{k}
 \threeJ{j}{j'}{l'}{m}{m'}{k'}={\delta_{l}^{l'}\delta_{k}^{k'}\over
2l+1} ,
 \label{eb4}
\end{equation}
where it is assumed that  the triangular condition is satisfied,
to obtain
\begin{equation}
 \threeJ{j}{j'}{L}{k}{-k}{0}
 z_L^M=
 (-1)^{-k}\threeJ{j}{j'}{0}{0}{0}{0} C_J
 \delta_{j}^{j'}\delta_{L}^0\delta_{M}^{0}.
 \label{eb5}
\end{equation}
Let us consider the possible cases in this equation separately.
For $L\not=0$, (\ref{eb5}) is simply
\begin{equation}
 \threeJ{j}{j'}{L}{k}{-k}{0} z_L^M=0,\qquad \forall j,j'\leq J.
 \label{eb6}
\end{equation}
The variables $z_L^M$  must be zero for $L=1,2,\dots,2J$, since
the 3-j symbols are non-vanishing. The other case, i.e., $L=0$,
does not give further information about $z_L^M$, since the
corresponding condition is trivially satisfied because of the
properties of the 3-j symbols~\cite{edmonds}. This completes the
proof of the equivalence between~(\ref{eb1f}) and~(\ref{eb2}).

{}From~(\ref{eb2}), and working along the same lines as Derka~{\em
et.~al.}~\cite{derka}, one can produce an algorithm for finite
POVMs. Suppose $J$ is integer (if it is not, consider the nearest
integer $\hat{J}>J$). We define $2J+1$ angles
$\phi_{s}=2s\pi/(2J+1)$; $s=0,1,\dots,2J$. Then $
\sum_{s=0}^{2J}Y^{M}_{L}(\theta,\phi_{s})=0$ ($M> 0$) for any
$\theta$, and we only need to solve
\begin{equation}
\label{eqzz} \sum_{k} c_{k} P_{L}(\cos{\theta_{k}})=0;\qquad
L=1,2,\dots,2J,
\end{equation}
where $P_{L}$ is the Legendre polynomial of degree~$L$. We choose
$\theta_{k}$ to be the $2J+2$ angles $\theta_{k}=k\pi/(2J+1)$;
$k=0,1,\dots,2J+1$; and define $c_{0}=c_{2J+1}=1$. Then, the
system~(\ref{eqzz}) of linear equations for $c_{1}$,
$c_{2}$,\dots, $c_{2J}$ always has a positive solution. Actually,
$c_{k}>1$ for $k=1,2,\dots,2J$. To summarize, the unit vectors
$\vec n_{r}\to\vec n_{k s}=(\sin\theta_{k}\cos\phi_{s},
\sin\theta_{k}\sin\phi_{s},\cos\theta_{k})$, along with the
corresponding weights $c_{r}\to c_{k s}\equiv c_{k}$ are
isotropically distributed, i.e., (\ref{eb1f}) is satisfied.

The above algorithm enables us to discretize any optimal
continuous POVM. Just take the very same state(s) $\ket{B}$ used
to generate the projectors $O(\vec n)$ and consider the new
(finite) set $O(\vec n_{r})= U(\vec n_{r})
\ket{B}\bra{B}U^\dagger(\vec n_{r})$. Modulo a trivial global
normalization factor, $\{O(\vec n_{r})\}$ defines a finite POVM.
The finite measurement thus obtained leads to the same fidelity
as the continuous one we started with. Moreover, since the
conditions~(\ref{eb2}) are exactly those used in~\cite{lpt} to
obtain minimal POVMs, it is clear that this construction also
provides minimal POVMs for general~$\ket{B}$ states. For
instance, the minimal POVM for $N=2$ has four outcomes pointing
to the vertices of a tetrahedron, while for $N=3$ there are six
outcomes corresponding to the vertices of an octahedron.

Finally, we would like to note that, as far as the fidelity is
concerned, Alice could also simulate a continuous isotropic
distribution of directions by using a finite set $\{\vec n_r\}$ of
isotropically distributed vectors (\ref{eb1f}) with \textit{a
priori} probability given by the weights $\{c_r/C_J\}$. The
fidelity will not change provided $J\ge(2j+1)/2$, where $j$ is
the total spin of the signal state ($j=N/2$ for a system of $N$
spins). For instance, if  $N=2$ and Alice uses unit  vectors
pointing to the vertices of an octahedron ($J=3/2$) with equal
probability $1/6$, the maximal fidelities  will be precisely
those shown in Table~I  for a truly (continuous) isotropic
distribution, namely, $F_P=3/4$ and $F_A=(3+\sqrt{3})/6$.

In summary, product
states of antiparallel spins represent an excellent balance
between feasibility of construction and capability to communicate
spin directions. For small number of spins their maximal fidelity
is remarkably close to the maximal value that can be possibly
achieved. For large $N$ these states lead to an average fidelity
that approaches unity faster than states with parallel spins,
although they do not exhibit the quadratic improvement of the
optimal states. We have thus proven that the truly optimal encoding
necessarily
requires entanglement. We have also obtained a simple set
of conditions for constructing finite measurements. These
conditions work for any eigenstate of the total spin  and,
therefore, also holds for product states.

We thank R. Tarrach and A. Brey for their collaboration in early
stages of this work, and M.~Lavelle for a careful reading of the
manuscript. Financial support from CICYT contract AEN99-0766 and
CIRIT contracts 1998SGR-00051, 1999SGR-00097 is acknowledged.

\end{document}